\documentstyle[prl,aps]{revtex}

\input epsf
\begin{document}
\draft
\twocolumn[\hsize\textwidth\columnwidth\hsize\csname@twocolumnfalse\endcsname

\title{Quasiparticle excitation in and around the vortex core of underdoped YBa$_2$Cu$_4$O$_8$ studied by site-selective NMR}
\author{ Kosuke Kakuyanagi$^{1}$, Ken-ichi Kumagai$^{1}$, and Yuji Matsuda$^{2}$}
\address{$^{1}$Division of Physics, Graduate School of Science, Hokkaido University, Kita-ku, Sapporo 060-0810, Japan}
\address{$^{2}$Institute for Solid State Physics, University of Tokyo, Kashiwanoha 5-1-5, Kashiwa, Chiba 277-8581, Japan}

\date{Received 12 April, 2001, Phys. Rev. B 65, 060503 (2002)}                                                                                                                                                                                                          

\maketitle

\begin{abstract}
We report a site-selective $^{17}$O spin-lattice relaxation rate  $T_1^{-1}$ in the vortex state of underdoped YBa$_2$Cu$_4$O$_8$.   We found that $T_1^{-1}$ at the planar sites exhibits an unusual nonmonotonic NMR frequency dependence.   In the region well outside the vortex core,  $T_1^{-1}$ cannot be simply explained by the density of states of the Doppler-shifted quasiparticles in the $d$-wave superconductor.   Based on  $T_1^{-1}$ in the vortex core region, we establish  strong evidence that the local density of states within the vortex core is strongly reduced.

\end{abstract}

]

\narrowtext


	The electronic structure in the vortex state is a fundamental problem in the physics of superconductors with unconventional symmetries.  Recent experimental results have revealed that the quasiparticle (QP) excitations in the vortex state of high-$T_c$ cuprates are very different from those of conventional $s$-wave superconductors.  Despite extensive studies on this subject,  the question of the QP excitations both inside and outside the vortex core is still far from being settled.  In fact,  STM experiments have revealed many unexpected properties in the spectrum of the vortex core \cite{stm}.   For example, a zero energy peak in the tunnel conductance due to the extended zero energy QP states, which is expected in the conventional theory of the $d$-wave superconductor,  is not observed at the center of the vortex core.  Although the novel electronic structure of the vortex core has been discussed in terms of several intriguing models, it remains controversial \cite{wang,franz2,himeda,arovas}.  On the other hand, well outside the vortex core, the QPs are considered to be in extended states due to the presence of gap nodes.  The dominant effect of these delocalized states in the magnetic field is the Doppler shift of the QP energy spectrum, which occurs due to the presence of a supercurrent flow around each vortex and generates a nonzero density of states (DOS) at the Fermi surface \cite{volovik}.  The $H$- and $T$-dependencies of the specific heat \cite{moler}, thermal conductivity \cite{taillefer}, and NMR Knight shift \cite{zhang} at low temperatures have generally been accepted as strong evidence for this effect.   However, rather surprisingly, neither the local DOS with a $1/r$ asymptotic tail expected from the Doppler shift nor the 4-fold symmetry of the extended QPs has been observed in the STM spectrum \cite{stm}.  Thus, a discrepancy still exists between the STM findings and the results of other experiments including thermodynamic measurement.  The situation therefore strongly confronts us with the need for a powerful probe of the QP excitation with spatial resolution.

	The measurements of the NMR spin-lattice relaxation rate $T_1^{-1}$ provide microscopic information on the low energy QP excitations in the superconducting state.   Recent experimental and theoretical studies have established that the frequency dependence of $T_1^{-1}$ measurements serves as a site selective probe for the electronic spectrum in the vortex state of the high-$T_c$ cuprates \cite{curro,takigawa,wortis,morr}.   Although the site-selective NMR, compared to STM, does not possess an atomic resolution , it has two advantages.  First, it is a {\it bulk probe, free from surface effects}.  Second, as pointed out by several authors \cite{franz,caxis}, the tunneling conductance obtained by STM is {\it not} simply related to the local DOS owing to the very anisotropic tunneling  matrix element along the $c$-axis.  The site selective NMR, on the other hand, especially at the oxygen sites in the plane where the antiferromagnetic spin fluctuation is cancelled, provides direct information on the local DOS \cite{masashi,bulut}.  Very recently,  the frequency dependence of $T_1^{-1}$ at the $^{17}$O sites (denoted as $^{17}T_1^{-1}$) in the vortex state of optimally doped YBa$_2$Cu$_3$O$_7$ has been reported \cite{curro}.  A key feature of the data is that $^{17}T_1^{-1}$ at the planar sites increases monotonically with NMR frequency well outside the vortex core, which is consistent with the delocalized QP DOS induced by the Doppler shift.   In this Letter, we attempted to measure the site-selective NMR in underdoped  YBa$_2$Cu$_4$O$_8$ extending the measurement to the vortex core region which has not yet been reported.    We have found an unusual NMR frequency dependence of $^{17}T_1^{-1}$ at the planar sites in the vortex lattice state, which is markedly different from that of YBa$_2$Cu$_3$O$_7$.  On the  basis of these results, we discuss the QP structure in and around the vortex core of YBa$_2$Cu$_4$O$_8$.

	High quality underdoped YBa$_2$Cu$_4$O$_8$ with $T_c$=81~K has been synthesized under a high-oxygen-pressure atmosphere.  Previous NMR studies have shown that the normal state spin fluctuations at the planar Cu and O sites for YBa$_2$Cu$_4$O$_8$ are similar in behavior to those for the 60~K phase of YBa$_2$Cu$_3$O$_{6+x}$ \cite{tomeno}.  The NMR data were obtained on an aligned powder of underdoped YBa$_2$Cu$_4$O$_8$,  which was isotopically enriched with $^{17}$O.  The $^{17}$O spin echo signals were obtained by a conventional pulse spectrometer under a field cooling condition (FCC) in a constant field ($H_0$=9.4~T) applied parallel to the $c$-axis by using a highly homogeneous superconducting magnet.  We stress here that the NMR measurements under the FCC are important because the Bean critical current associated with the sweeping magnetic field not only produces the field gradient in the crystal but also seriously influences $^{17}T_1^{-1}$ by producing an additional Doppler shift to the QPs.  

	The final NMR spectra were obtained by convolution of the respective FT-spectra of the spin echo signals measured by a step of 50~kHz.    Given that the spectra of the central transition of the chain O(1), the planar O(2,3), and the apical O(4) overlap at low temperature,  $^{17}T_1$ is determined by the recovery curves of the quadrupole split O(2,3) satellite.  The inset of Fig.~1(a) shows the planar $^{17}$O (O(2) and O(3)) spectrum of the satellite transition at 100~K.  The slightly-split between the satellite spectra of the oxygen sites of O(2) and O(3) is due to a small difference in electric field gradient. The procedure to determine $T_1$ is as follows. The spin echo intensities are measured as a function of time after saturation pulses. Then the FT spectrum of each echo is obtained.  Each frequency component of the FT spectra is fitted individually by $R(t)=(M(\infty)-M(t))/M(\infty)= 1/35\exp(-2Wt) + 3/56\exp(-6Wt) + 1/40\exp(-12Wt) + 25/56\exp(-20Wt) + 25/56\exp(-30Wt)$ for the (3/2$\leftrightarrow$1/2) transition. Here, we calculated $^{17}T_1$ (=1/2$W$) for each frequency point at the 3~kHz interval with a gaussian weight function of $\sigma$=3~kHz.

		   The local field profile in the vortex state is obtained by approximating $H_{loc}(\mbox{\boldmath$r$})$ with the London result, 
\begin{equation}
H_{loc}(\mbox{\boldmath$r$})=H_0\sum_{G}\exp{(-i\mbox{\boldmath$G\cdot
r$})}~~\frac{\exp({-\xi_{ab}^2\mbox{\boldmath$G$}^2/2})}{1+\mbox{\boldmath$G$}^2\lambda_{ab}^2},
\end{equation}
where $\mbox{\boldmath$G$}$ is a reciprocal vector of the vortex lattice, $\mid \mbox{\boldmath$r$}\mid$ is the distance from the center of the core, $\xi_{ab}$ is the in-plane coherence length, and $\lambda_{ab}$ is the in-plane penetration length.  The thin solid lines in Figs.~1 (a) and (b) depict the histogram at particular local fields which are given by the local field distribution $f(H_{loc})=\int_{\Omega} \delta [H_{loc} (\mbox{\boldmath$r$})-H_{loc}]d^2\mbox{\boldmath$r$}$, where $\Omega$ is the magnetic unit cell.  In the calculation we used $\xi$=15\AA~and $\lambda_{ab}=1600$~\AA~reported in Ref.\cite{pana} and assumed the square vortex lattice.  The magnetic field is the lowest at the center of the vortex square lattice (C-point in the inset of Fig.~1(b)), and it is the highest at the center of the vortex core (A-point).  The intensity of the histogram shows a peak at the field corresponding to the saddle point of the vortices (B-point).  Because the local NMR frequency is proportional to the local magnetic field, we are able to produce a spatial image of the low-lying QP excitation spectrum.  

	The solid lines in Figs.~1 (a) and (b) represent the NMR spectrum at the planar and the apical sites in the superconducting state, respectively.  A clear asymmetric pattern, which originates from the local field distribution associated with the Abrikosov vortex lattice (the

\begin{figure}
\centerline{\epsfxsize 8.6cm \epsfbox{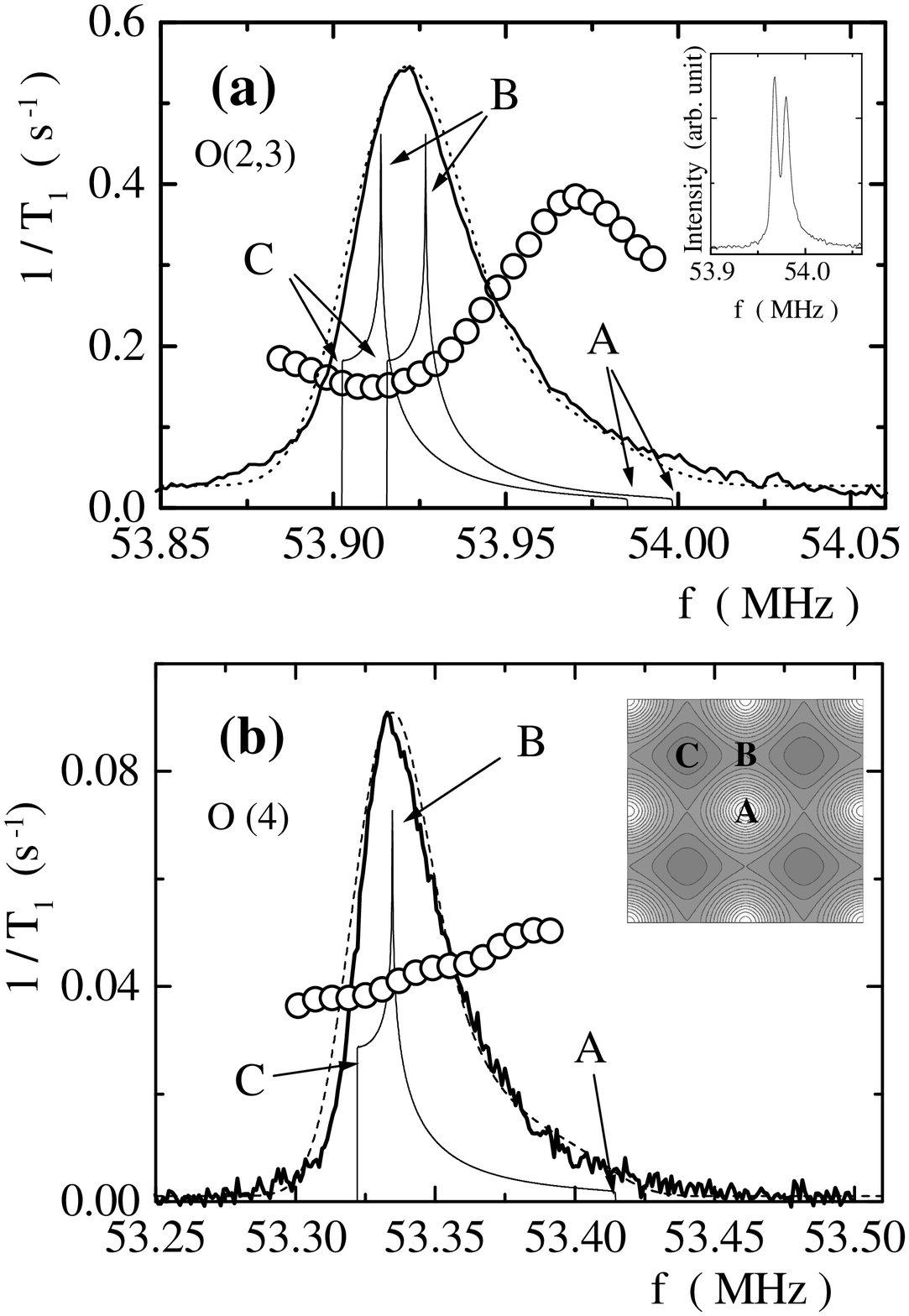}}
\caption{The planar $^{17}$O(2, 3) (a) and the apical $^{17}$O(4) (b) satellite (3/2$\leftrightarrow$1/2 transition) spectrum (solid lines) at $T$=20~K and 10~K, respectively.  The open circles show the $T_1^{-1}$ at corresponding sites.  The thin solid lines represent Redfield patterns for planar $^{17}$O(2), $^{17}$O(3) and apical $^{17}$O(4) sites. The broken lines represent the simulation spectrum convoluted with the Gaussian broadening.  For details, see the text.  The inset of Fig.1~(a) shows the satellite spectrum of $^{17}$O(2,3) at 100K. The inset of Fig.1~(b) shows the image of field distribution in the square lattice of vortices.  A-, B-, and C-points represent the vortex core center, the saddle point of the field distribution, and the middle of the vortex square lattice, respectively. }
\end{figure}
	
\noindent Redfield's	pattern), is observed well below $T_c$ in both sites.  The broken lines in Figs.~1 (a) and (b) depict the spectrum convoluted with the Gaussian broadening function, $f(H_{loc})=e^{-2H_{loc}^2/\sigma^2}$.  (The broken line in Fig.~1(a) is obtained by the sum of the spectrum at the O(2) and (3) sites.) 	Using $\sigma$=24~kHz, the theoretical curves well reproduce the data in both planar and apical sites.  The open circles in Figs.~1 (a) and (b) display the NMR frequency dependence of $^{17}T_1^{-1}$ at the planar (denoted as $^{17}T_{1,pl}^{-1}$)  and at the apical ($^{17}T_{1,ap}^{-1}$) sites, respectively.   A clear nonmonotonic frequency dependence is observed at the planar sites.   With increasing frequency,  $^{17}T_{1,pl}^{-1}$ first decreases,  shows a minimum near the B-point, then increases.  Finally, at high  frequency when approaching 
	
\begin{figure}
\centerline{\epsfxsize 8.2cm \epsfbox{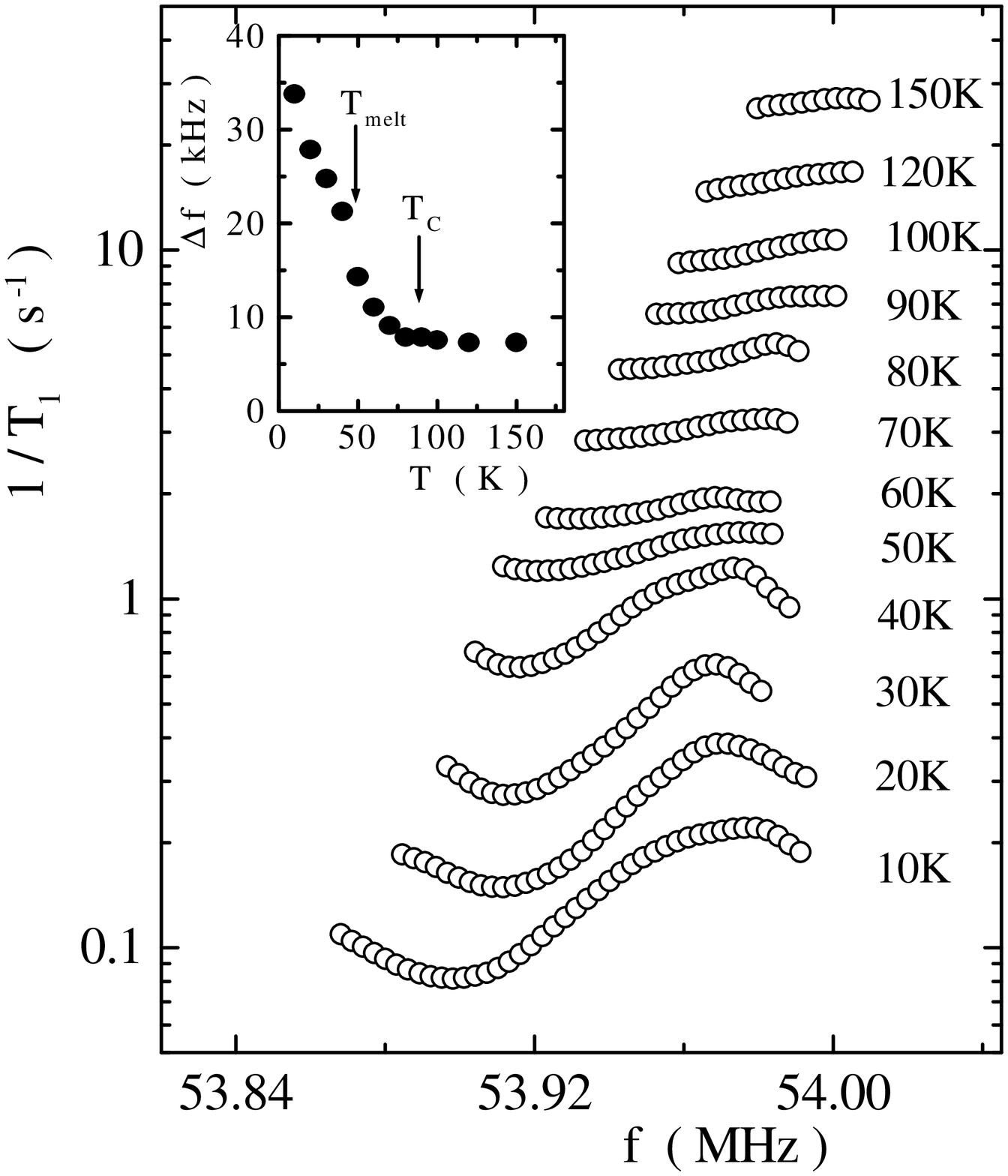}}
\caption{1/$T_1$ of $^{17}$O(2, 3) at planar sites, $^{17}T_{1,pl}^{-1}$, as a function of frequency for various temperatures.  The inset shows the half width at half intensity of the spectrum of $^{17}$O(2) as a function of temperature. $T_c$ is the superconducting transition temperature, and $T_{melt}$ is the vortex lattice melting temperature. \protect \cite{qui}}
\end{figure}

\noindent the vortex core regime (A-point),  $^{17}T_{1,pl}^{-1}$  again decreases.  This frequency dependence is in contrast to the  result of Ref.\cite{curro}, in which $^{17}T_{1,pl}^{-1}$ shows  monotonic  increases with frequency.    In contrast to $^{17}T_{1,pl}^{-1}$,  $^{17}T_{1,ap}^{-1}$ shows little frequency dependence. 

     Figure 2 shows the frequency dependence of $^{17}T_{1,pl}^{-1}$ at various temperatures. The inset shows the half width at half intensity of the spectrum of $^{17}$O(2) as a function of temperature. One can clearly see a slight increase of the line width below $T_c$ and the large broadening due to the vortex lattice below $T_{melt}\sim45$~K.\cite{qui} $^{17}T_{1,pl}^{-1}$ becomes nearly independent of frequency at $\sim45$~K, above which the vortex lattice melts into disordered vortex liquid. The frequency dependences of $^{17}T_{1,pl}^{-1}$ are remarkable below $\sim45$~K. This result reinforces the fact that the observed frequency dependence of $^{17}T_{1,pl}^{-1}$ at lower temperatures originates from the field distribution associated with the vortex lattice. Figure 3 displays the $T$-dependencies of $^{17}T_{1,pl}^{-1}$ and $^{17}T_{1,ap}^{-1}$.  For the planar sites, we show the largest and the smallest values of $^{17}T_{1,pl}^{-1}$ for corresponding temperatures. The $T$-dependence of the 
	 
\begin{figure}
\centerline{\epsfxsize 7.9cm \epsfbox{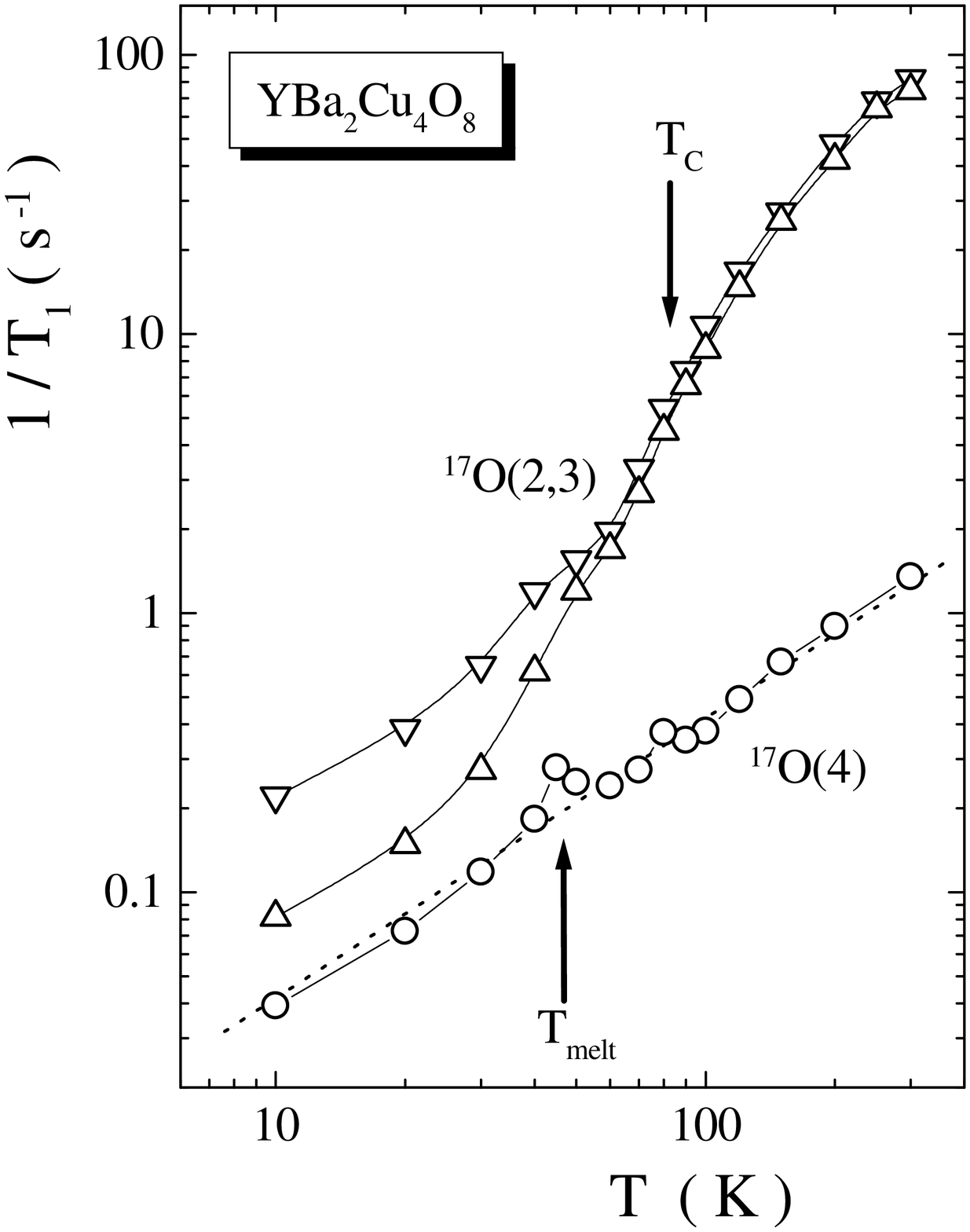}}
\caption{$T$-dependences of 1/$T_1$ of planar $^{17}$O(2,3) and apical $^{17}$O(4) sites. $\bigtriangledown$ ($\bigtriangleup$) represent the largest (the smallest) values of $^{17}T_{1,pl}^{-1}$ of $^{17}$O(2,3) across frequencies. The dotted line for the $^{17}$O(4) data represents the $T_1T$=const. relation.}
\end{figure}
 
\noindent smallest values is very similar to results reported previously. \cite{tomeno} Because $T_1$ is determined from averaged recovery curves over all frequency components by a conventional method, $T_1$ is attributed to the value from the most dominant components of the signals, namely, in the superconducting region far from vortex cores (around the B point).   With decreasing $T$, $^{17}T_{1,pl}^{-1}$ decreases faster than $T$  above $T_c$ due to the pseudogap, and does not show an anomaly at $T_c$ \cite{masashi}.  On the other hand,  $1/^{17}T_{1,ap}$ is nearly  proportional to $T$ except at $\simeq$45~K, where a peak in  $^{17}T_{1,pl}^{-1}$ associated with the vortex lattice melting is clearly observed.\cite{curro}

	We first discuss the relaxation mechanism.  It should be noted that the spin diffusion effect from the core region is unimportant in the present case because of the large broadening of the spectra below $T_M$ \cite{curro}.  Generally, $^{17}T_1^{-1}$ is composed of a contribution from vortex vibration $^{17}T_{1,VV}^{-1}$ and a contribution from the QP scattering $^{17}T_{1,QP}^{-1}$ \cite{curro,bula}:
\begin{equation}
\frac{1}{^{17}T_1}=\frac{1}{^{17}T_{1,VV}}+\frac{1}{^{17}T_{1,QP}}.
\end{equation}
The vortex vibration influences $^{17}T_1$ through the transverse fluctuation of the local magnetic field.  It has been suggested that the vortex vibration plays an important role in $T_1$ for very anisotropic Tl$_2$Ba$_2$CuO$_6$ \cite{bula}.   We consider that in YBa$_2$Cu$_4$O$_8$, the QP scattering predominates $^{17}T_1$ at the planar sites, while the vortex vibration seemingly influences $^{17}T_1$ at the apical sites, based on the following explanation.   First, as shown in Figs.~1 (a) and (b),  $^{17}T_{1,ap}^{-1}$ shows much smaller frequency dependence than $^{17}T_{1,pl}^{-1}$.   Second,  if the contribution from the vortex vibration dominates, then  $^{17}T_{1,ap}^{-1}$ should be the same as $^{17}T_{1,pl}^{-1}$.  However, as shown in Fig.~3,  $^{17}T_{1,ap}^{-1}$ are smaller than $^{17}T_{1,pl}^{-1}$ by a factor of $\sim 5$.   Third, the vortex lattice melting seriously influences  $^{17}T_{1,ap}^{-1}$, while its influence on $^{17}T_{1,pl}^{-1}$ is fairly small.\cite{curro}
  
	Having established that the $^{17}T_{1,pl}^{-1}$ is dominated by the QP excitation, the next question is the frequency dependence.   We first discuss $^{17}T_1^{-1}$ in the low frequency region well outside the core.  Within the framework of the semi-classical approximation, it has been pointed out that the local DOS $N(E=0,r)$ well outside the core is dominantly produced by the Doppler shift of the delocalized QP spectrum,   $E(\mbox{\boldmath $p$})\rightarrow E(\mbox{\boldmath $p$})-\mbox{\boldmath $v$}_s \cdot \mbox{\boldmath $p$}$, in the presence of the supercurrent flow with velocity $\mid \mbox{\boldmath $v$}_s\mid=\hbar/2mr$.   As a result,  there is a finite $N(E=0,r)(\propto \mid \mbox{\boldmath $v$}_s\mid)$, which decays as $1/r$ for a single vortex and varies 
slightly more modestly than $1/r$ in the vortex lattice state.   According to Ref.\cite{wortis}, this local DOS gives rise to a monotonous increase of $^{17}T_{1,pl}^{-1}$ with frequency. 

In the intermediate frequency regime where  $^{17}T_{1,pl}^{-1}$ increases,  the Doppler shift appears to be important for the local DOS.   However, {\it the observed decrease of $^{17}T_{1,pl}^{-1}$ when going from C- to B-point is apparently inconsistent with the Doppler shift}.    There are two possible origins for this unusual behavior in $^{17}T_{1,pl}^{-1}$.   One is the competition between the thermally excited QPs and the Doppler-shifted  QPs in the presence of the strong antiferromagnetic fluctuation which is strongly suppressed by the Doppler shift, as pointed out in Ref.\cite{morr}.  The other is the vortex lattice effect,  which is not taken into account in the semi-classical theory of Doppler shift.  According to the numerical calculation, the localized QPs in the Andreev bound state around the vortex core extend to $(\pm \pi, \pm \pi)$ directions, and can overlap the QPs of the neighboring vortices \cite{ichioka}.  This effect can increase the QP DOS at the C-point, depending on the vortex lattice structure.  At present it is not clear which effect is relevant to $^{17}T_{1,pl}^{-1}$ at low frequency.    The fact that the low frequency behavior of  $^{17}T_{1,pl}^{-1}$ cannot be simply explained by the Doppler shifted QPs implies some key features of the QP excitation outside the vortex core. 

	 Finally we discuss $^{17}T_{1,pl}^{-1}$ at high frequencies, at which $^{17}T_{1,pl}^{-1}$ begins to decrease after reaching maximum.  It is apparent that this behavior is again incompatible with the Doppler shift.   As pointed out in Ref.\cite{franz}, the semi-classical description of the QPs breaks down in the vicinity of the core region ($r<2\xi$) where the superconducting order parameter varies rapidly.  Therefore it is natural to consider that the reduction of $^{17}T_{1,pl}^{-1}$ occurs in the vortex core regime.  It should be noted that the antiferromagnetic fluctuation, if it exists, always increases $^{17}T_{1,pl}^{-1}$.   These considerations lead us conclude that  {\it the decrease of $^{17}T_{1,pl}^{-1}$ is caused by the suppression of the  QP DOS within the vortex core.}   This is the first experimental evidence of the reduction of the local  DOS within the vortex core, which was probed by a {\it bulk measurement}.  In the pure $d$-wave superconductors, the local DOS at the center of the vortex core should be enhanced, as it is in  $s$-wave superconductors, owing to the interference of the  Andreef scattered QPs at the both sides of the pair potential.    It would follow that  $^{17}T_{1,pl}^{-1}$ should be enhanced in the vortex core regime, which is apparently inconsistent with the present results.   The present results are seemingly consistent with the STM measurements, which show that only one QP energy level is present within the core for the optimally doped YBa$_2$Cu$_3$O$_7$ and Bi$_2$Sr$_2$CaCu$_2$O$_{8+\delta}$ \cite{stm}. Several possible origins for the reduction of DOS have been  proposed as follows: (1) different symmetry, such as $d_{x^2-y^2}+id_{xy}$, is induced within the core \cite{franz2};  (2) the Andreef bound state  is split due to the large $\Delta^2/\varepsilon_F$ value where $\Delta$ is the superconducting gap and $\varepsilon_F$ is the Fermi energy;  (3) the strong antiferromagnetic interaction changes the core state dramatically, as discussed by the calculation based on the $t-J$ model \cite{himeda} and the SO(5) theory \cite{arovas}.  To clarify the origin of the reduction of the local DOS within the vortex core, more detailed information of the QP structure, especially in the underdoped regime, is strongly required.

	In summary, we measured the site-selective spin-lattice NMR relaxation rate in and around the vortex core of underdoped YBa$_2$Cu$_4$O$_8$.   Nonmonotonic NMR frequency dependence of $T_{1,pl}^{-1}$ indicates that $T_{1,pl}^{-1}$ well outside the vortex core regime cannot be simply explained by the Doppler-shifted delocalized quasiparticles of the $d$-wave superconductor and that the local DOS in the vortex core regime is strongly reduced.

We thank T.~Isobe and S.~Shamoto for assisting with the crystal preparation.  We also acknowledge helpful discussions with  M.~Ichioka, Y.~Kato, K.~Machida, K.~Maki, D.K.~Morr, M.~Ogata, M.~Takigawa, A.~Tanaka, and I.~Vekhter.  This work has been supported by a Grant-in-Aid from Ministry of Education, Culture, Sports, Science and Technology.

{\it Note added in proof.} ~After submission (at April 12, 2001) of this manuscript, we are aware of an article (in Nature 413 (October 4, 2001) 501) of similar NMR study on the spatial dependence of $T_1^{-1}$ in near-optimally doped YBa$_2$Cu$_3$O$_7$.

\end{document}